\documentclass[pdflatex,sn-basic]{sn-jnl}

\usepackage{graphicx}%
\usepackage{wrapfig, multirow, bigdelim, bm, bbm, float}%
\usepackage{amssymb, amsmath, amsfonts, ascmac, color, mathtools}%
\usepackage{latexsym, pifont, longtable, fancybox, subcaption}%
\usepackage{amsthm}%
\usepackage{mathrsfs}%
\usepackage[title]{appendix}%
\usepackage{xcolor}%
\usepackage{textcomp}%
\usepackage{manyfoot}%
\usepackage{booktabs}%
\usepackage{url, appendix, algorithm}
\usepackage{algorithmicx}%
\usepackage{algpseudocode}%
\usepackage{listings}%
\usepackage{tikz}
\usepackage{import}
\usepackage{comment}
\usepackage{tabularx}
\usepackage[normalem]{ulem}

\theoremstyle{thmstyleone}%
\newtheorem{theorem}{Theorem}
\theoremstyle{thmstyletwo}%

\theoremstyle{thmstylethree}%
\newtheorem{definition}{Definition}%

\begin{document}

\title[Generalized Poisson Matrix Factorization]{Generalized Poisson Matrix Factorization for Overdispersed Count Data}


\author[1]{\fnm{Ryo} \sur{Ohashi}}\email{ryotennis0807@gmail.com}

\author*[2]{\fnm{Hiroyasu} \sur{Abe}}\email{hiro00012@gmail.com}
\equalcont{These authors contributed equally to this work.}

\author[3]{\fnm{Fumitake} \sur{Sakaori}}\email{sfumitake001e@g.chuo-u.ac.jp}
\equalcont{These authors contributed equally to this work.}

\affil[1]{\orgdiv{Graduate School of Science and Engineering}, \orgname{Chuo University}, \orgaddress{\street{1-13-27 Kasuga}, \city{Bunkyo-ku}, \postcode{112-8551}, \state{Tokyo}, \country{Japan}}}

\affil*[2]{\orgdiv{School of Pharmaceutical Sciences}, \orgname{Wakayama Medical University}, \orgaddress{\street{25-1 shichibancho}, \city{Wakayama-shi}, \postcode{640-8156}, \state{Wakayama}, \country{Japan}}}

\affil[3]{\orgdiv{Faculty of Science and Engineering}, \orgname{Chuo University}, \orgaddress{\street{1-13-27 Kasuga}, \city{Bunkyo-ku}, \postcode{112-8551}, \state{Tokyo}, \country{Japan}}}


\abstract{
Non-negative matrix factorization (NMF) is widely used as a feature extraction technique for matrices with non-negative entries, such as image data, purchase histories, and other types of count data. In NMF, a non-negative matrix is decomposed into the product of two non-negative matrices, and the approximation accuracy is evaluated by a loss function. If the Kullback–Leibler divergence is chosen as the loss function, the estimation coincides with maximum likelihood under the assumption that the data entries are distributed according to a Poisson distribution. To address overdispersion, negative binomial matrix factorization has recently been proposed as an extension of the Poisson-based model. However, the negative binomial distribution often generates an excessive number of zeros, which limits its expressive capacity. In this study, we propose a non-negative matrix factorization based on the generalized Poisson distribution, which can flexibly accommodate overdispersion, and we introduce a maximum likelihood approach for parameter estimation. This methodology provides a more versatile framework than existing models, thereby extending the applicability of NMF to a broader class of count data.}




\maketitle

\section{Introduction}\label{sec1}
Many types of real-world data, such as pixel intensities in images, word counts in text, or purchase frequencies in transaction records, are inherently non-negative.
For such data, it is often desirable that any low-rank representation also consist of non-negative components, so that the decomposed factors can be interpreted in terms of the original features.
Traditional linear decomposition methods, such as Singular Value Decomposition (SVD), represent a data matrix as a combination of orthogonal basis vectors, but the resulting components typically include both positive and negative values.
When applied to non-negative data, this discrepancy requires ad hoc adjustments such as truncating negative entries or rescaling the data, which alters the basis vectors and makes interpretation difficult.
To address this issue, Non-negative Matrix Factorization (NMF) was proposed as a decomposition method that enforces non-negativity in the factorized matrices.

NMF decomposes a non-negative data matrix into the product of two non-negative matrices.
\citet{Lee1999} introduced efficient algorithms for obtaining locally optimal solutions under non-negativity constraints via the auxiliary function method.
They considered Euclidean distance (squared error) and Kullback–Leibler (KL) divergence as loss functions to quantify the discrepancy between the observed matrix and its factorized approximation.
Since their seminal work, numerous NMF variants based on alternative divergence measures have been proposed.
Many of these approaches are equivalent to maximum likelihood estimation (MLE) of the factorization parameters under specific probabilistic assumptions: NMF based on Euclidean distance corresponds to MLE under a Normal assumption, whereas NMF based on KL divergence corresponds to MLE under a Poisson assumption.
Thus, NMF can naturally be formulated within a statistical modeling framework.

Because NMF is defined for non-negative data, several variants have been developed specifically for count data matrices.
As noted earlier, the KL divergence–based NMF corresponds to a Poisson model, which serves as a natural starting point for extensions to count data.
Subsequent research has proposed variants tailored to additional distributional features.
For instance, zero-inflated Poisson matrix factorization (ZIPMF) \citep{simchowitz2013zero} models the data with a zero-inflated Poisson distribution, which assigns a higher probability to zeros than the standard Poisson.
This formulation is particularly useful for sparse datasets, such as purchase records, because it distinguishes between incidental non-purchases and a complete lack of purchase intent.
Negative binomial matrix factorization (NBMF) \citep{Gouvert2020} instead employs the negative binomial distribution, which incorporates a dispersion parameter and thereby accounts for overdispersion.
This feature makes NBMF suitable for applications such as music listening data, where user-specific preferences influence play counts.
Owing to these modeling capabilities, ZIPMF and NBMF have been widely applied to recommendation systems based on user–item matrices.
Further extensions have also been proposed, including zero-inflated Tweedie models \citep{Abe2017} and zero-inflated negative binomial models \citep{Abe2018, Abe2019}.
Since all of these approaches are formulated within a statistical modeling framework, they can be naturally extended to Bayesian settings.
For instance, \citet{cemgil2009bayesian} reformulated KL divergence–based NMF as a Bayesian hierarchical model, known as Bayesian NMF.

The choice of probability distribution assumed for the data matrix in NMF critically influences approximation accuracy and is therefore central to model design. NBMF, which employs the negative binomial distribution, is commonly used for overdispersed count data. The negative binomial distribution, however, assigns an increasingly high probability to zero counts as dispersion grows \citep{MINAMI2007210}, making it less suitable when overdispersion coexists with a substantial proportion of non-zero entries. To address this limitation, we propose a generalized Poisson matrix factorization (GPMF). The generalized Poisson distribution \citep{Consul01111973} incorporates a dispersion parameter yet, at comparable dispersion levels, tends to yield fewer zeros and exhibits heavier tails than the negative binomial distribution. Building on this distributional choice, we introduce a maximum-likelihood estimation procedure that jointly estimates the basis and coefficient matrices as well as the dispersion parameter, thereby extending the applicability of NMF to a broader class of overdispersed count data.

The remainder of this paper is organized as follows.
Section \ref{Sec: Negative Binomial Matrix Factorization} reviews the basics of NMF.
Section \ref{Sec: Related Works} summarizes NMF approaches for count data, including Poisson matrix factorization (PMF), its zero-inflated extension (ZIPMF), and NBMF.
Section \ref{Sec: Generalized Poisson Matrix Factorization} introduces the proposed generalized Poisson matrix factorization (GPMF), defines the generalized Poisson distribution, formulates the model, and derives maximum-likelihood update rules.
Section \ref{Sec: Experimental Result} reports simulation studies comparing the proposed method with PMF and NBMF in terms of approximation accuracy.
Finally, Section \ref{Sec: Conclusion} concludes with a summary and directions for future research.

\section{Non-negative Matrix Factorization}\label{Sec: Non-negative Matrix Factorization}
Non-negative matrix factorization (NMF) is a matrix decomposition method that imposes non-negativity constraints on both the observed data and the factor matrices.  
Let $Y \in \mathbb{R}_{\geq 0}^{I \times J}$ denote a non-negative matrix of dimension $I \times J$.  
NMF approximates $Y$ by the product of two non-negative matrices
\begin{align}
	Y \simeq S = WH^{\top},
\end{align}
where $W \in \mathbb{R}_{\ge 0}^{I \times K}$ and $H \in \mathbb{R}_{\ge 0}^{J \times K}$ are called the basis and coefficient matrices, respectively.  
Each entry $y_{ij}$ of $Y$ is then approximated by
\begin{align}
	y_{ij} \simeq s_{ij} = \sum_{k=1}^{K} w_{ik} h_{jk}. \label{eq: equation of factorization}
\end{align}
The rank $K$, the number of columns of $W$ and $H$, is chosen so that $K \ll \min\{I,J\}$.

Although the factorization (\ref{eq: equation of factorization}) is typically formulated as minimizing the elementwise divergence between $y_{ij}$ and $s_{ij}$, it can also be interpreted as maximum likelihood estimation under a probability model for $y_{ij}$ with expectation $s_{ij}$,
\begin{align}
	y_{ij} \sim P(s_{ij},\theta), \label{eq: statistical model representation of NMF}
\end{align}
where $P(s_{ij},\theta)$ denotes a probability distribution with mean $s_{ij}$ and additional parameters $\theta$.

Imposing non-negativity on both the observed elements and the factorized components offers two main advantages.  
First, because real-world data are often non-negative, the resulting factorized matrices are more readily interpretable.  
Second, since each observation is represented as a sum of non-negative components, the non-negativity constraint induces a tendency toward sparsity in the coefficient matrix.  
This sparsity facilitates the extraction of essential information from the basis matrix for each observation.

\section{Related Works}\label{Sec: Related Works}
Various extensions of NMF have been proposed for count data, each adopting a different probability distribution to capture specific characteristics of the observations.  
In what follows, we review representative approaches, beginning with Poisson matrix factorization (PMF) as the simplest formulation, followed by its zero-inflated and negative binomial extensions, and concluding with Bayesian formulations.

\subsection{Poisson Matrix Factorization (PMF)}
Poisson matrix factorization (PMF) is an NMF method based on the Poisson distribution as a probability model,
\begin{align*}
	y_{ij} \sim \mathrm{Poi}(s_{ij}),
\end{align*}
with $\mathrm{Poi}(\mu)$ denoting the Poisson distribution and having probability mass function
\begin{align*}
	f(x) = \frac{\mu^{x} e^{-\mu}}{x!}. \label{eq: pmf of Poisson}
\end{align*}
PMF was originally formulated by \citet{Lee1999} as the problem of minimizing the Kullback–Leibler (KL) divergence, equivalent to maximum likelihood estimation as described above.  
Using the auxiliary function method (Theorem \ref{Auxiliary Function Method} in Appendix \ref{Sec: Auxiliary Function Method}), they derived an efficient iterative algorithm for estimating the factorized matrices.

\subsection{Zero-Inflated Poisson Matrix Factorization (ZIPMF)}
Zero-inflated Poisson matrix factorization (ZIPMF) \citep{simchowitz2013zero} is an NMF method based on the zero-inflated Poisson (ZIP) distribution,
\begin{align}
	y_{ij} \sim \mathrm{ZIP}(\pi, s_{ij}),
\end{align}
with $\mathrm{ZIP}(\pi,\lambda)$ denoting a discrete distribution whose probability mass function is
\begin{align}
	f(x) = \pi \mathbbm{1}_{x = 0} + (1-\pi)\frac{e^{-\lambda}\lambda^{x}}{x!}.
\end{align}
The ZIP is a mixture of a degenerate distribution at zero, $\mathbbm{1}_{k=0}$, and a Poisson distribution with parameter $\lambda \in \mathbb{R}_{>0}$, combined with mixing proportion $\pi \in (0,1)$.  
Compared with the standard Poisson distribution, the ZIP assigns greater probability to zero.

ZIPMF is particularly effective for datasets in which many entries are zero and these zeros have heterogeneous interpretations.  
For instance, in product purchase count data, each user does not buy every available product, leading to many zero entries.  
Some of these zeros arise by chance, whereas others reflect a complete lack of purchase intent.  
Conventional NMF cannot distinguish between these two cases.  
In contrast, ZIPMF models chance zeros as generated by the Poisson component and structural zeros as generated by the degenerate component, thereby enabling extraction of user-specific characteristics.

\subsection{Negative Binomial Matrix Factorization (NBMF)}\label{Sec: Negative Binomial Matrix Factorization}
Negative binomial matrix factorization (NBMF) \citep{Gouvert2020} is an NMF method based on the negative binomial (NB) distribution,
\begin{align}
	y_{ij} \sim \mathrm{NB}\left(\alpha,\  \frac{s_{ij}}{s_{ij} + \alpha} \right), \label{eq: NBMF}
\end{align}
with $\mathrm{NB}(r,p)$ denoting a discrete distribution whose probability mass function is
\begin{align}
	f(x) = \frac{\Gamma(x+r)}{x!\Gamma(r)}p^{x}(1-p)^{r},
\end{align}
where $r \in \mathbb{R}_{>0}$ is the dispersion parameter and $p \in [0,1]$ is the success probability.  

Count data, such as product purchases or music listening histories, often exhibit overdispersion, characterized by variability larger than that expected under a Poisson model.  
In such cases, applying Poisson matrix factorization (PMF) can lead to poor approximation accuracy.  
To address this issue, \citet{Gouvert2020} proposed NBMF and demonstrated that it outperforms PMF in analyzing user preferences based on music listening data.  
They also derived a divergence criterion associated with their model \eqref{eq: NBMF}.  
Furthermore, NBMF can be extended to a zero-inflated version, analogous to ZIPMF, known as zero-inflated negative binomial matrix factorization (ZINBMF), which was proposed by \citet{Abe2018,Abe2019}.

\subsection{Bayesian formulation for NMF}
Bayesian formulations of NMF provide a principled framework for incorporating prior information and quantifying uncertainty in the factorization process.  
\citet{cemgil2009bayesian} introduced a hierarchical model in which $K$ latent variables---rather than the observed variable $y_{ij}$ itself---are modeled as Poisson variables whose sum equals the observed entry.  
This model reduces to PMF when the latent variables are marginalized out and can therefore be regarded as a Bayesian generalization of PMF.  
Building on this framework, Bayesian extensions have been developed for ZIPMF \citep{simchowitz2013zero}, NBMF \citep{Gouvert2018}, and ZINBMF \citep{Abe2018}.

\section{Generalized Poisson Matrix Factorization (GPMF)}\label{Sec: Generalized Poisson Matrix Factorization}
As reviewed in Section \ref{Sec: Negative Binomial Matrix Factorization}, negative binomial matrix factorization addresses overdispersed count data. The negative binomial distribution assigns increasingly high probability to zero counts as the dispersion parameter $r$ decreases, that is, as overdispersion increases. Consequently, NBMF can be ill-suited when overdispersion occurs without pronounced zero inflation, particularly in settings with a substantial proportion of nonzero entries. To overcome this limitation, we propose generalized Poisson matrix factorization, an NMF framework built on the generalized Poisson distribution. Like the negative binomial, the generalized Poisson includes a dispersion parameter, yet at comparable dispersion levels it places less mass at zero and exhibits heavier tails. We also derive a maximum likelihood estimation procedure that jointly estimates the basis and coefficient matrices together with the dispersion parameter.

\subsection{Generalized Poisson (GP) distribution}
The Poisson distribution, in which the mean and variance coincide, is a common generative model for count data. In practice, however, data often exhibit discrepancies between the observed mean and variance even when a Poisson assumption is made. To address this limitation, \citet{Consul01111973} introduced the generalized Poisson distribution as an extension of the standard Poisson. The GP includes a parameter $\eta$, analogous to the Poisson rate, together with an additional parameter $\mu$ that controls dispersion. When $\mu$ is positive, the variance exceeds the mean; when $\mu$ is zero, the variance equals the mean; and when $\mu$ is negative, the variance is smaller than the mean. \citet{yusuf2015performance} showed that in regression tasks with overdispersed count data, GP regression yields better predictive accuracy than classical Poisson regression. These findings indicate that the GP is particularly well-suited to modeling overdispersed count data when $\mu$ takes positive values.

A generalized Poisson random variable with parameters $\eta \in \mathbb{R}_{>0}$ and $-1 < \mu < 1$ is denoted as $Y \sim \mathrm{GP}(\eta, \mu)$, whose probability mass function, expectation, and variance are given by
\begin{align}
	f(x) &= \eta \frac{(\eta + \mu x)^{x-1} e^{-(\eta + \mu x)}}{x!}, \quad x \ge 0, \label{mass_GP}\\
	E[Y] &= \frac{\eta}{1-\mu}, \quad V[Y] = \frac{\eta}{(1-\mu)^3}. \label{eq: expectation and variance}
\end{align}
When $\mu = 0$, these expressions reduce to the probability mass function, mean, and variance of the standard Poisson distribution \eqref{eq: pmf of Poisson}, showing that the generalized Poisson includes the Poisson as a special case. For $\mu < 0$ the distribution represents underdispersion, whereas for $\mu > 0$ it captures overdispersion. In this study we focus on the overdispersed case with $\mu > 0$.

\begin{figure}[t]
	\centering
	\includegraphics[keepaspectratio, width=0.5\linewidth]{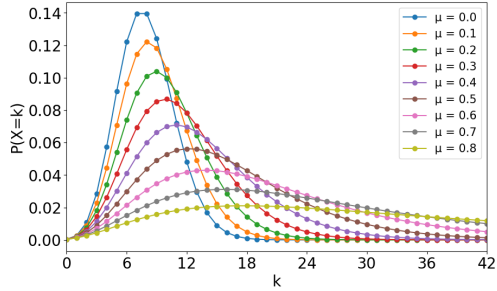}
	\caption{Graphs of the probability mass function of GP with varied $\mu$ and fixed $\eta=8.0$.}
	\label{fig: GP}
\end{figure}

Figure \ref{fig: GP} illustrates the probability mass function of the GP for different values of $\mu$ with $\eta = 8.0$ fixed. As seen from Figure \ref{fig: GP} and equation \eqref{eq: expectation and variance}, increasing $\mu$ shifts the peak of the distribution to the right and makes it more gradual, thereby increasing both the mean and the variance.

For application to NMF, we adopt a reparameterized form of the GP introduced by \citet{Consul01111973}, given by
\begin{align}
	Y \sim \text{GP}\left(\frac{\lambda}{1+\theta}, \frac{\theta}{1+\theta}\right), \label{eq: form of GP for NMF}
\end{align}
with the correspondence
\begin{align}
	\eta = \lambda(1-\mu), \quad \mu = \frac{\theta}{1+\theta}, \label{eq: reparameterization of GP}
\end{align}
where $\lambda \in \mathbb{R}_{\ge0}$ and $\theta \in \mathbb{R}_{\ge0}$. Under this parameterization, the expectation and variance are expressed as
\begin{align}
	E[Y] = \lambda, \quad V[Y] = \lambda(1+\theta)^2,
\end{align}
which shows that $\lambda$ determines the mean, while $\theta$ controls the degree of dispersion.

\begin{figure}[t]
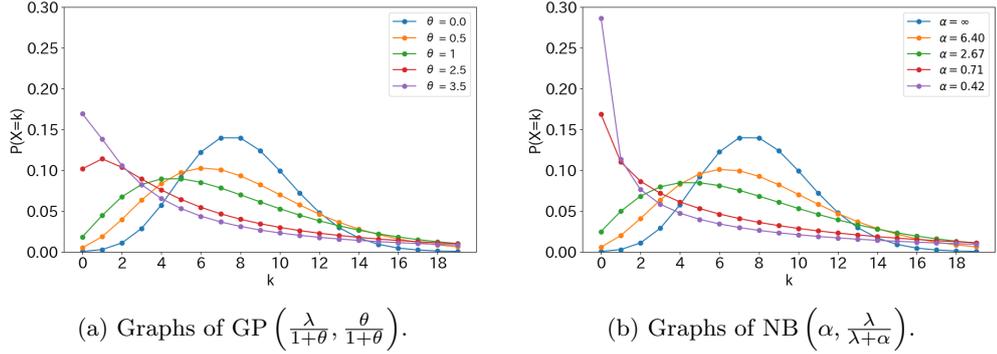

	\begin{minipage}[b]{0.48\linewidth}
		\centering
		\includegraphics[keepaspectratio, width=\linewidth]{GPMF_PMF.pdf}
		\subcaption{Graphs of $\text{GP}\left(\frac{\lambda}{1+\theta},\frac{\theta}{1+\theta}\right)$.}
	\end{minipage}
	\hspace{0.03\columnwidth}
	\begin{minipage}[b]{0.48\linewidth}
		\centering
		\includegraphics[keepaspectratio, width=\linewidth]{NBMF_PMF.pdf}
		\subcaption{Graphs of $\mathrm{NB}\left(\alpha,\frac{\lambda}{\lambda+\alpha}\right)$.}
	\end{minipage}
	\caption{GP and NB with same mean and variance. 
		$\lambda=8$ is fixed and $\alpha$ is tuning according to $\theta$.}
	\label{fig: GP and NB}
\end{figure}

Figure \ref{fig: GP and NB}(a) illustrates the GP defined in equation \eqref{eq: form of GP for NMF} with different values of $\theta$, showing that the peak of the distribution shifts gradually to the left as $\theta$ increases. This behavior indicates that $\theta$ plays a role analogous to the dispersion parameter $\alpha$ in the negative binomial distribution. Figure \ref{fig: GP and NB}(b) depicts a negative binomial distribution with the same mean and variance as the GPs in Figure \ref{fig: GP and NB}(a). To match the GP variance $\lambda(1+\theta)^2$, the dispersion parameter of the NB is set to $\alpha=\frac{\lambda}{\theta^{2}+2\theta}$. Comparing the two distributions for $\theta=2.5$, $\alpha=0.71$, and for $\theta=3.5$, $\alpha=0.42$, shows that the probability $P[x=0]$ is consistently lower under the GP than under the NB. Moreover, as $\alpha$ decreases and the variance grows, the NB places sharply increasing mass at zero, whereas the GP does not.
As shown in Appendix \ref{Sec: Kurtosis of GP}, the GP has greater kurtosis than the NB, reflecting its heavier tails. When the mean and variance are matched, this property leads the GP to allocate less probability to zero while distributing more mass in the tails. As the variance grows, the NB places an increasingly large probability at zero, whereas the GP increases more moderately at zero and instead develops progressively heavier right tails. These tendencies are consistent with the findings of \citet{Joe2005}, who showed that the GP assigns lower probability to zero and exhibits heavier tails than the NB under the same mean and variance. Appendix~\ref{Sec: Kurtosis of GP} further confirms this behavior by analytically demonstrating that the GP has greater kurtosis than the NB.
Taken together, these features indicate that the GP can serve as a more suitable model than the NB for over-dispersed data where zeros are not overwhelmingly dominant.

\subsection{Model and Loss Function of GPMF}
We now introduce the proposed NMF framework based on the generalized Poisson distribution. 
Each observed entry $y_{ij}$ is assumed to be independently generated from a generalized Poisson distribution with mean $s_{ij}>0$ and variance $s_{ij}(1+\theta_{i})^{2}$,
\begin{align*}
	y_{ij}\sim\mathrm{GP}\left(\frac{s_{ij}}{1+\theta_{i}},\frac{\theta_{i}}{1+\theta_{i}}\right),
\end{align*}
where $s_{ij}=\sum_{k=1}^{K}w_{ik}h_{jk}$.
When $\theta_{i}=0$ for all $i=1,2,\dots,I$, the formulation reduces to PMF, whereas positive values of $\theta_{i}$ allow the model to account for overdispersion. 

In the GPMF setting, the dispersion parameter $\theta$ is assumed to vary across rows. 
This reflects the idea that, in data such as product purchase counts or music listening frequencies, items may share similar means yet exhibit different levels of variability. 
Accordingly, treating items as variables (rows of the matrix) and users as random samples, the model specifies a row-specific dispersion parameter $\theta_{i}$. 
Although the exposition here assumes heterogeneity across $i$, an equivalent formulation can be obtained by letting the parameter vary with $j$ instead, by relabeling indices.

Parameter estimation is carried out by minimizing the negative log-likelihood
\begin{align}
	-\ell(Y|W,H,\Theta) &= \sum_{i=1}^{I}\sum_{j=1}^{J}-\ell(y_{ij}|s_{ij},\theta_{i}) \nonumber \\[6pt]
	& = \sum_{i=1}^{I}\sum_{j=1}^{J}\left[-\log\frac{s_{ij}}{1+\theta_{i}} -(y_{ij}-1)\log\left(\frac{s_{ij}+\theta_{i} y_{ij}}{1+\theta_{i}}\right)\right. \nonumber \\[-4pt]
	&\qquad\qquad\qquad\qquad\qquad\qquad\qquad\left. + \frac{s_{ij} + \theta_{i} y_{ij}}{1+\theta_{i}}+\log(y_{ij}!) \right], \label{eq: negative loglikelihood of GPMF}
\end{align}
where $\Theta=(\theta_{1},\ldots,\theta_{I})$ with $\theta_{i}\in\mathbb{R}_{>0}$. 
The estimation task thus consists of determining $W$, $H$, and $\Theta$, which can be achieved through an iterative updating algorithm that monotonically decreases the objective function \eqref{eq: negative loglikelihood of GPMF}.

\subsection{Update equations of parameters in GPMF}
We now derive the update equations for the parameters in GPMF.  
Starting from the loss function in \eqref{eq: negative loglikelihood of GPMF}, we collect the terms that do not depend on $W$, $H$, or $\Theta$ into a constant.  
This yields
\begin{align}
	-\ell(Y|W,H,\Theta)
	&= -\sum_{i=1}^{I}\sum_{j=1}^{J}\log s_{ij}
	   + \sum_{i,j:y_{ij}=0}\log s_{ij}
	   - \sum_{i,j:y_{ij}\ge 1}(y_{ij}-1)\log\!\left(s_{ij}+\theta_i y_{ij}\right) \notag\\
	&\quad + \sum_{i=1}^{I}\sum_{j=1}^{J}y_{ij}\log(1+\theta_i)
	   + \sum_{i=1}^{I}\sum_{j=1}^{J}\frac{s_{ij}+\theta_i y_{ij}}{1+\theta_i}
	   + \mathrm{const} \notag\\
	&= -\sum_{i,j:y_{ij}\ge 1}\Big\{\log s_{ij}+(y_{ij}-1)\log(s_{ij}+\theta_i y_{ij})\Big\} \notag\\
	&\quad + \sum_{i=1}^{I}\sum_{j=1}^{J}y_{ij}\log(1+\theta_i)
	   + \sum_{i=1}^{I}\sum_{j=1}^{J}\frac{s_{ij}+\theta_i y_{ij}}{1+\theta_i}
	   + \mathrm{const.}
\end{align}
This reformulation separates the components of the objective function in a way that will be useful for applying auxiliary-function techniques in the subsequent derivations.

For the term $-(y_{ij}-1)\log(s_{ij}+\theta_i y_{ij})$ in the loss function,  
we apply Theorem \ref{Jensen's inequality} to obtain the upper bound
\begin{align}
	-(y_{ij}-1)\log(s_{ij}+\theta_i y_{ij})
	\le -(y_{ij}-1)\Bigg\{\eta_{ij1}\log\!\left(\frac{s_{ij}}{\eta_{ij1}}\right)
	+ \eta_{ij2}\log\!\left(\frac{\theta_i y_{ij}}{\eta_{ij2}}\right)\Bigg\},
\end{align}
where $\eta_{ij1},\eta_{ij2}>0$ with $\eta_{ij1}+\eta_{ij2}=1$.  
The advantage of this bound is that the right-hand side decomposes into a sum of functions depending separately on $s_{ij}$ and $\theta_i$.  
Moreover, the bound becomes tight when minimized with respect to $\eta_{ij1}$ and $\eta_{ij2}$, which occurs at
\begin{align*}
	\eta_{ij1} = \frac{s_{ij}}{s_{ij}+\theta_i y_{ij}}, 
	\qquad
	\eta_{ij2} = \frac{\theta_i y_{ij}}{s_{ij}+\theta_i y_{ij}}.
\end{align*}

Next, we apply Theorem \ref{Jensen's inequality} to the term 
\[
	-\left(1+(y_{ij}-1)\eta_{ij1}\right)\log s_{ij},
\]
which yields
\begin{align}
	-\left(1+(y_{ij}-1)\eta_{ij1}\right)\log s_{ij}
	\le -\left(1+(y_{ij}-1)\eta_{ij1}\right)
	\sum_{k=1}^{K}\lambda_{ijk}\log\!\left(\frac{w_{ik}h_{jk}}{\lambda_{ijk}}\right),
\end{align}
where $\lambda_{ijk}>0$ and $\sum_{k=1}^{K}\lambda_{ijk}=1$.  
Equality holds when
\[
	\lambda_{ijk}=\frac{w_{ik}h_{jk}}{\sum_{k=1}^{K}w_{ik}h_{jk}}.
\]
This decomposition allows the dependence on $s_{ij}$ to be expressed as a weighted sum of the factors $w_{ik}h_{jk}$, which will be essential for deriving element-wise update rules.

Combining the two applications of Jensen's inequality, the loss function in \eqref{eq: negative loglikelihood of GPMF} can be bounded above by an auxiliary function.  
Specifically, substituting the inequalities for 
$-(y_{ij}-1)\log(s_{ij}+\theta_i y_{ij})$ and 
$-\bigl(1+(y_{ij}-1)\eta_{ij1}\bigr)\log s_{ij}$, 
we obtain
\begin{align}
	G_{\mathrm{GPMF}}&(W,H,\Theta,\lambda,\eta) \nonumber \\
	=& \sum_{i,j:y_{ij}\ge1}\Bigl\{-(1+(y_{ij}-1)\eta_{ij1})\sum_{k=1}^{K}\lambda_{ijk}\log(w_{ik}h_{jk})\Bigr\} \nonumber \\
	&\hspace*{2em}-\left\{\sum_{i,j:y_{ij}\ge1}(y_{ij}-1)\eta_{ij2}\right\}\log\theta_i
	+\sum_{i=1}^{I}\sum_{j=1}^{J}y_{ij}\log(1+\theta_i) \nonumber \\
	&\hspace*{2em}+\sum_{i=1}^{I}\sum_{j=1}^{J}\frac{s_{ij}+\theta_i y_{ij}}{1+\theta_i}
	+ \mathrm{const.} \label{eq: auxiliary function of GPMF}
\end{align}
Here, the variables $\lambda=(\lambda_{ijk})$ and $\eta=(\eta_{ij1},\eta_{ij2})$ are introduced through Jensen’s inequality, and the right-hand side is separable with respect to $W$, $H$, and $\Theta$.  
This separability enables element-wise optimization and forms the basis of the iterative update scheme.

With the auxiliary function in \eqref{eq: auxiliary function of GPMF} at hand, we minimize it under non-negativity constraints, leading to the inequality-constrained problem
\begin{align}
	&\underset{W,H,\Theta}{\mathrm{minimize}}\quad G_{\mathrm{GPMF}}(W,H,\Theta,\lambda,\eta) \nonumber\\
	&\mathrm{subject\ to}\quad w_{ik}\ge0,\quad h_{jk}\ge0,\quad \theta_i\ge0 \nonumber\\
	&\qquad\qquad\qquad\qquad (i=1, 2, \dots,I;\; j=1, 2, \dots,J;\; k=1, 2, \dots,K), \label{min_GPMF}
\end{align}
where $\lambda=(\lambda_{ijk})$ and $\eta=(\eta_{ij1},\eta_{ij2})$ collect the Jensen weights introduced above. Because $G_{\mathrm{GPMF}}$ is separable across matrix elements, the problem decomposes into simple one-variable subproblems, which we handle via a Lagrangian formulation in the next step.

In this case, the optimization problem \eqref{min_GPMF} can be reformulated as an unconstrained extremum problem through the Lagrangian function
\begin{align*}
	\mathcal{L}_{\mathrm{GPMF}}&(W,H,\Theta) \\
	&= G_{\mathrm{GPMF}}(W,H,\Theta,\lambda,\eta)
	- \left\{ 
		\sum_{i=1}^{I}\sum_{k=1}^{K}a_{ik}w_{ik} 
		+ \sum_{j=1}^{J}\sum_{k=1}^{K}b_{jk}h_{jk} 
		+ \sum_{i=1}^{I}c_{i}\theta_{i} 
	\right\}.
\end{align*}
Here, $a_{ik}$, $b_{jk}$, and $c_{i}$ are the Lagrange multipliers associated with the non-negativity constraints.  
According to the Karush--Kuhn--Tucker (KKT) conditions, the necessary conditions for a local minimum require that, for all $i,j,k$,
\begin{align*}
	&\frac{\partial \mathcal{L}_{\mathrm{GPMF}}}{\partial w_{ik}} = 0,\quad 
	\frac{\partial \mathcal{L}_{\mathrm{GPMF}}}{\partial h_{jk}} = 0,\quad 
	\frac{\partial \mathcal{L}_{\mathrm{GPMF}}}{\partial \theta_{i}} = 0,\\
	&w_{ik}\ge0,\quad h_{jk}\ge0,\quad \theta_{i}\ge0,\\
	&a_{ik}\ge0,\quad b_{jk}\ge 0,\quad c_{i}\ge 0,\\
	&a_{ik}w_{ik}=0,\quad b_{jk}h_{jk}=0,\quad c_{i}\theta_{i}=0.
\end{align*}

By noting that $w_{ik}>0$ and $h_{jk}>0$, we solve 
$\tfrac{\partial \mathcal{L}_{\mathrm{GPMF}}}{\partial w_{ik}}=0$ 
and $\tfrac{\partial \mathcal{L}_{\mathrm{GPMF}}}{\partial h_{jk}}=0$ 
with respect to $w_{ik}$ and $h_{jk}$, respectively. 
This leads to
\begin{align*}
	-\frac{1}{w_{ik}}\sum_{j:y_{ij}\ge 1}\!\left\{(1+(y_{ij}-1)\eta_{ij1})\lambda_{ijk}\right\}
	+ \sum_{j=1}^{J}\frac{h_{jk}}{1+\theta_{i}} &= 0,\\
	-\frac{1}{h_{jk}}\sum_{i:y_{ij}\ge 1}\!\left\{(1+(y_{ij}-1)\eta_{ij1})\lambda_{ijk}\right\}
	+ \sum_{i=1}^{I}\frac{w_{ik}}{1+\theta_{i}} &= 0.
\end{align*}
Rearranging gives
\begin{align*}
	w_{ik} &= \frac{\sum_{j:y_{ij}\ge 1}(1+(y_{ij}-1)\eta_{ij1})\lambda_{ijk}}{\sum_{j=1}^{J}\tfrac{h_{jk}}{1+\theta_{i}}},\\[6pt]
	h_{jk} &= \frac{\sum_{i:y_{ij}\ge 1}(1+(y_{ij}-1)\eta_{ij1})\lambda_{ijk}}{\sum_{i=1}^{I}\tfrac{w_{ik}}{1+\theta_{i}}}.
\end{align*}
Substituting 
$\lambda_{ijk}=\tfrac{w_{ik}h_{jk}}{s_{ij}}$ and 
$\eta_{ij1}=\tfrac{s_{ij}}{s_{ij}+\theta_{i}y_{ij}}$, 
we obtain the multiplicative update rules
\begin{align}
	w_{ik} &\leftarrow 
	w_{ik}\frac{\sum_{j=1}^{J}\!\left(y_{ij}\tfrac{s_{ij}+\theta_{i}}{s_{ij}+\theta_{i}y_{ij}}\cdot\tfrac{h_{jk}}{s_{ij}}\right)}{\sum_{j=1}^{J}\tfrac{h_{jk}}{1+\theta_{i}}},
	\quad
	h_{jk} \leftarrow 
	h_{jk}\frac{\sum_{i=1}^{I}\!\left(y_{ij}\tfrac{s_{ij}+\theta_{i}}{s_{ij}+\theta_{i}y_{ij}}\cdot\tfrac{w_{ik}}{s_{ij}}\right)}{\sum_{i=1}^{I}\tfrac{w_{ik}}{1+\theta_{i}}}.
	\label{update_GPMF}
\end{align}

Next, noting that $\theta_{i}>0$, we solve 
$\tfrac{\partial \mathcal{L}_{\mathrm{GPMF2}}}{\partial \theta_{i}}=0$ 
with respect to $\theta_{i}$:
\begin{align}
	-&\frac{1}{\theta_{i}}\left\{\sum_{j:y_{ij}\ge 1}(y_{ij}-1)\eta_{ij2}\right\}
	+ \frac{1}{1+\theta_{i}}\left(\sum_{j:y_{ij}\ge 1}y_{ij}\right) \\
	&\qquad + \frac{1}{(1+\theta_{i})^{2}}\left\{\sum_{j=1}^{J}(y_{ij}-s_{ij})\right\} 
	- c_{i} = 0 \notag\\
	&\Leftrightarrow\quad\left\{ J - n_{0} + \sum_{j:y_{ij}\ge1}(y_{ij}-1)\eta_{ij1} \right\}\theta_{i}^{2}\\
	&\qquad +\left\{\sum_{j=1}^{J}(y_{ij}-s_{ij})-\sum_{j:y_{ij}\ge1}(y_{ij}-1)\eta_{ij2}
	+ J - n_{0} + \sum_{j:y_{ij}\ge1}(y_{ij}-1)\eta_{ij1} \right\}\theta_{i} \notag\\
	&\qquad - \left\{ \sum_{j:y_{ij}\ge1}(y_{ij}-1)\eta_{ij2} \right\} = 0.
	\label{GPMF_theta_eq}
\end{align}

Here, we define $n_{0}=\sum_{j:y_{ij}=0}1$. Introducing
\begin{align*}
	\alpha_{i} &= J-n_{0}+\sum_{j:y_{ij}\ge1}(y_{ij}-1)\eta_{ij1}, \\
	\beta_{i}  &= \left\{\sum_{j=1}^{J}(y_{ij}-s_{ij})\right\}-\gamma_{i}+\alpha_{i}, \\
	\gamma_{i} &= \sum_{j:y_{ij}\ge1}(y_{ij}-1)\eta_{ij2},
\end{align*}
equation \eqref{GPMF_theta_eq} reduces to a quadratic equation in $\theta_{i}$:
\begin{align}
	\alpha_{i}\theta_{i}^{2} + \beta_{i}\theta_{i} - \gamma_{i} = 0,\quad \theta_{i}>0.
\end{align}

From the quadratic formula, the update rule for $\theta_{i}$ is obtained as
\begin{align}
	\theta_{i} \leftarrow \frac{-\beta_{i}+\sqrt{\beta_{i}^{2}+4\alpha_{i}\gamma_{i}}}{2\alpha_{i}}.
	\label{update_GPMF_theta}
\end{align}

\begin{algorithm}[b]
	\caption{Maximum Likelihood Estimation of $W$ and $H$ in GPMF}         
	\label{algGPMF}
	\begin{algorithmic}
		\Require $W,H,\Theta$
		\Ensure $W,H,\Theta$
		\While{the change in the log-likelihood is larger than a tolerance}
		\For{$i = 1$ to $I$, $j = 1$ to $J$, $k = 1$ to $K$}
		\State $w_{ik} \leftarrow w_{ik}\frac{\sum_{j=1}^{J}\left(y_{ij}\frac{s_{ij}+\theta_{i}}{s_{ij} + \theta_{i}y_{ij}}\cdot\frac{h_{jk}}{s_{ij}}\right)}{\sum_{j=1}^{J}\frac{h_{jk}}{1+\theta_{i}}}$
		\State $h_{jk} \leftarrow h_{jk}\frac{\sum_{i=1}^{I}\left(y_{ij}\frac{s_{ij}+\theta_{i}}{s_{ij} + \theta_{i} y_{ij}}\cdot\frac{w_{ik}}{s_{ij}}\right)}{\sum_{i=1}^{I}\frac{w_{ik}}{1+\theta_{i}}}$
		\EndFor
		\State $\alpha_{i}\leftarrow J-n_{0}+\sum_{j:y_{ij}\ge1}(y_{ij}-1)\frac{s_{ij}}{s_{ij} + \theta_{i} y_{ij}}$
		\State $\gamma_{i}\leftarrow \sum_{j:y_{ij}\ge1}(y_{ij}-1)\frac{\theta_{i} y_{ij}}{s_{ij} + \theta_{i} y_{ij}}$
		\State $\beta_{i}\leftarrow \left\{\sum_{j=1}^{J}(y_{ij}-s_{ij})\right\}-\gamma_{i}+\alpha_{i}$
		\State $\theta_{i}\leftarrow \frac{-\beta_{i}+\sqrt{\beta_{i}^{2}+4\alpha_{i}\gamma_{i}}}{2\alpha_{i}}$
		\EndWhile
	\end{algorithmic}
\end{algorithm}

Since we assume $w_{ik},h_{jk},\theta_{i}>0$, the corresponding Lagrange multipliers vanish, i.e., $a_{ik}=b_{jk}=c_{i}=0$.  
The algorithm for GPMF based on these update rules is summarized in Algorithm \ref{algGPMF}.

The update equations \eqref{update_GPMF} and \eqref{update_GPMF_theta} correspond to the case where the dispersion parameter of the GP distribution varies across rows, i.e., $\theta_{i}$.  
In the case where the parameter is assumed to be common across all elements, i.e., $\theta$, the update equations are derived in the same way. They take the following form:
\begin{align}
	w_{ik} &\leftarrow w_{ik}(1+\theta)\frac{\sum_{j=1}^{J}\left(y_{ij}\frac{s_{ij}+\theta}{s_{ij} + \theta y_{ij}}\cdot\frac{h_{jk}}{s_{ij}}\right)}{\sum_{j=1}^{J}h_{jk}},\\
	h_{jk} &\leftarrow h_{jk}(1+\theta)\frac{\sum_{i=1}^{I}\left(y_{ij}\frac{s_{ij}+\theta}{s_{ij} + \theta y_{ij}}\cdot\frac{w_{ik}}{s_{ij}}\right)}{\sum_{i=1}^{I}w_{ik}}, 
	\label{update_GPMF1} \\
	\theta &\leftarrow \frac{-\beta+\sqrt{\beta^{2}+4\alpha\gamma}}{2\alpha},
	\label{update_GPMF1_theta}
\end{align}
where
\begin{align*}
	\alpha &= IJ-n_{0}+\sum_{i,j:y_{ij}\ge1}(y_{ij}-1)\eta_{ij1},\\
	\beta  &= \left\{\sum_{i=1}^{I}\sum_{j=1}^{J}(y_{ij}-s_{ij})\right\}-\gamma+\alpha,\\
	\gamma &= \sum_{i,j:y_{ij}\ge1}(y_{ij}-1)\eta_{ij2}.
\end{align*}

\section{Experimental Result}\label{Sec: Experimental Result}
In this section, we evaluate the reconstruction accuracy of PMF, NBMF, and our GPMF using synthetic data. 

The size of the data matrix is set to $I=50$ and $J=100$. The factorization rank, i.e., the number of columns in the two factor matrices, is fixed at $K=5$. Each element of the true basis and coefficient matrices, $W_{0}\in\mathbb{R}_{\ge0}^{I\times K}$ and $H_{0}\in\mathbb{R}_{\ge0}^{J\times K}$, is generated independently from $\mathrm{Gamma}(1.5,1.5)$, where $\mathrm{Gamma}(\theta,\lambda)$ denotes the Gamma distribution with shape $\theta$ and rate $\lambda$. Using these matrices, we compute $S_{0}=W_{0}H_{0}^{\top}$, and then generate the data matrix $Y$ elementwise from a GP as
\[
y_{ij}\sim\mathrm{GP}\left(\dfrac{(S_0)_{ij}}{1+\theta_{0,i}},\dfrac{\theta_{0,i}}{1+\theta_{0,i}}\right).
\]

We examine the reconstruction accuracy under five levels of dispersion, $\theta_{0,i}=0,\,0.5,\,1,\,1.5$, and $2$, where the dispersion parameter is constant within each row. In addition, we consider a heterogeneous setting in which $\theta_{0,i}$ varies across rows, with the values $0,\,0.5,\,1,\,1.5$, and $2$ assigned to exactly the same number of rows. For each setting, $100$ synthetic datasets are generated, and the mean squared errors (MSEs) of $W$, $H$, and $S$ are computed.  

We apply PMF, NBMF, and GPMF to each synthetic dataset. For NBMF, the dispersion parameter is fixed at $\alpha=5$. For all algorithms, the stopping criterion is set to $\epsilon = 10^{-6}$, where convergence is determined based on the normalized difference in log-likelihoods. Specifically, the absolute difference between successive log-likelihood values is divided by the absolute value of the updated log-likelihood plus one. This criterion behaves as an absolute tolerance when the log-likelihood is small and as a relative tolerance when it is large, providing a hybrid stopping rule.

For each algorithm, two types of initializations for the factors $w_{ik}$ and $h_{jk}$ were considered. The first is an NNDSVD-based initialization, which is commonly used in NMF \citep{boutsidis2008nndsvd}. To avoid excessive sparsity in the initial factors, we employed a slight modification of the standard NNDSVD: the first singular component was initialized using absolute values of the corresponding singular vectors, which improves numerical stability under multiplicative updates. The second is a random initialization. Specifically, the elements $w_{ik}$ and $h_{jk}$ were independently generated from a $\mathrm{Gamma}(1,1)$ distribution and subsequently rescaled so that the sample mean of the initial intensity $S_{0,ij}$ coincides with the sample mean of the observed data $y_{ij}$. For GPMF, the dispersion parameters $\theta_i$ were initialized at 1 for all $i$.

To ensure a fair comparison, we account for the indeterminacy in scaling and column permutation inherent in matrix factorization before computing the MSEs. Specifically, the columns of $H$ are normalized to unit norm, and the corresponding columns of $W$ are rescaled so that the product $WH^{\top}$ remains unchanged. Finally, the columns of the estimated factor matrices are permuted to best align with those of the ground truth.  

\begin{table}[b]
	\centering
	\caption{The means of 100 $\mathrm{MSE}_{W_{0}}$, $\mathrm{MSE}_{H_{0}}$, and $\mathrm{MSE}_{S_{0}}$.}\label{tab: MSEs}
	\begin{tabular}{ccrrrrrr}
		\toprule
		 & & \multicolumn{2}{c}{PMF} & \multicolumn{2}{c}{NBMF} & \multicolumn{2}{c}{GPMF}\\
		\cmidrule(lr){3-4} \cmidrule(lr){5-6} \cmidrule(lr){7-8}
		$\theta_0$ & MSE & NNDSVD & random & NNDSVD & random & NNDSVD & random \\
		\midrule       
			& $W_{0}$ & 57.1 & 29.0 & 83.6 & 37.5 & 52.7 & \textbf{28.6} \\
		0	& $H_{0}$ & 0.00155 & 0.00117 & 0.00192 & 0.00152 & 0.00152 & \textbf{0.00116} \\
			& $S_{0}$ & 0.991 & \textbf{0.956} & 1.13 & 1.10 & 0.993 & 0.957 \\		
		\midrule
			& $W_{0}$ & 98.0 & \textbf{85.0} & 114 & 102 & 110 & 90.1 \\
		0.5	& $H_{0}$ & 0.00297 & \textbf{0.00277} & 0.00335 & 0.00332 & 0.00311 & 0.00288 \\
			& $S_{0}$ & 2.57 & 2.52 & 2.78 & 2.78 & 2.59 & \textbf{2.51} \\		
		\midrule
			& $W_{0}$ & 163 & \textbf{156} & 177 & 166 & 186 & 158 \\
		1.0	& $H_{0}$ & 0.00428 & \textbf{0.00412} & 0.00465 & 0.00458 & 0.00444 & 0.00429 \\
			& $S_{0}$ & 4.80 & 4.73 & 5.15 & 5.01 & 4.59 & \textbf{4.25} \\		
		\midrule
			& $W_{0}$ & 239 & 234 & 258 & 243 & 264 & \textbf{211} \\
		1.5	& $H_{0}$ & 0.00529 & \textbf{0.00519} & 0.00565 & 0.00559 & 0.00556 & 0.00526 \\
			& $S_{0}$ & 7.52 & 7.45 & 8.20 & 7.79 & 6.89 & \textbf{5.94} \\	
		\midrule
			& $W_{0}$ & 327 & 318 & 356 & 332 & 355 & \textbf{251} \\
		2.0	& $H_{0}$ & 0.00615 & 0.00602 & 0.00645 & 0.00634 & 0.00652 & \textbf{0.00597} \\
			& $S_{0}$ & 10.9 & 10.8 & 12.3 & 11.5 & 9.95 & \textbf{7.61} \\		
		\midrule
			& $W_{0}$ & 236 & 216 & 263 & 216 & 287 & \textbf{148} \\
		heterogeneous	& $H_{0}$ & 0.00542 & 0.00523 & 0.00566 & 0.00561 & 0.00538 & \textbf{0.00392} \\
			& $S_{0}$ & 6.71 & 6.56 & 7.73 & 6.66 & 6.24 & \textbf{3.98} \\		
		\bottomrule
	\end{tabular}
\end{table}

The results are summarized in Table \ref{tab: MSEs}. Across all settings, the choice of initialization was found to have a non-negligible impact on performance. In particular, random initialization with multiple starts consistently achieved smaller MSEs than the NNDSVD-based initialization for all three methods. This tendency was especially pronounced for GPMF, suggesting that multi-start random initialization is more effective in exploring the complex likelihood surface induced by dispersion parameters.

Regarding model comparison, the relative performance of PMF, NBMF, and GPMF strongly depended on the degree of overdispersion in the data. When little or no overdispersion was present, PMF tended to perform favorably and often achieved reconstruction errors comparable to, or smaller than, those of the more flexible models. This result is consistent with the fact that PMF directly matches the Poisson assumption underlying the data-generating process in such settings, rendering additional flexibility unnecessary. As the degree of overdispersion increased, however, the advantage of PMF diminished and GPMF became increasingly superior. In particular, when overdispersion varied across rows, GPMF consistently outperformed both PMF and NBMF. These results demonstrate the benefit of explicitly modeling heterogeneous dispersion through row-specific parameters, which allows GPMF to adapt to structural variability that cannot be captured by PMF or NBMF.

The performance of NBMF depends strongly on the choice of the dispersion-related parameter $\alpha$. Although appropriate tuning of $\alpha$ can improve reconstruction accuracy in certain scenarios, selecting a suitable value in practice is nontrivial and often requires additional heuristics or prior knowledge. In contrast, GPMF estimates dispersion parameters directly from the data, thereby avoiding manual specification and providing a more robust and adaptive modeling framework in the presence of strong or heterogeneous overdispersion.

Overall, these simulation results indicate that PMF is preferable when the Poisson assumption is approximately valid, whereas GPMF offers clear advantages as overdispersion becomes stronger or more heterogeneous. This behavior highlights the importance of explicitly accounting for dispersion structures in probabilistic matrix factorization, particularly in applications involving count data with substantial variability beyond the Poisson assumption.

Although random initialization with multiple starts generally yields better performance, it is computationally more demanding. From a practical perspective, NNDSVD-based initialization provides a convenient alternative, and GPMF initialized by NNDSVD may still offer a reasonable compromise between computational efficiency and reconstruction accuracy in situations where computational resources are limited. 

\section{Conclusion}\label{Sec: Conclusion}
In this study, we proposed a novel nonnegative matrix factorization method, termed GPMF, for modeling overdispersed count data with many non-zero entries. The proposed method assumes that observations follow a generalized Poisson distribution and provides an efficient algorithm that simultaneously estimates dispersion-controlling parameters from the data. Through simulation studies using synthetic data, we demonstrated that GPMF achieves lower reconstruction errors than conventional approaches when overdispersion is present, particularly in settings with strong or heterogeneous dispersion.

At the same time, our results indicate that simpler models such as PMF remain preferable when the Poisson assumption is approximately valid. Compared with NBMF, whose performance depends sensitively on the choice of dispersion-related parameters, GPMF avoids manual tuning by estimating dispersion parameters directly, thereby offering a more robust and adaptive modeling framework. These properties make GPMF a practical alternative for count data analysis in situations where the degree of overdispersion is unknown or varies across observations.

Future work includes extending the proposed framework to a Bayesian formulation, which would enable uncertainty quantification through posterior inference, such as credible intervals for latent factors and reconstructed intensities. Such an extension would allow for more principled assessment of estimation uncertainty and predictive variability in applications involving noisy or highly overdispersed count data.

\begin{appendices}

\section{Auxiliary Function Method}\label{Sec: Auxiliary Function Method}

The optimization of NMF is generally formulated as a constrained problem with nonnegativity restrictions, which precludes analytical solutions.  
\citet{Lee1999} introduced an efficient algorithm based on the \emph{auxiliary function method}, which iteratively minimizes an upper bound of the objective function rather than the function itself.  
This approach guarantees monotonic convergence of the objective value.  
For completeness, we briefly summarize the definitions and key results used in this study.

\begin{definition}\label{Auxiliary Function}
Let $\theta = \{\theta_i\}_{1\leq i \leq I}$ denote a set of variables, and let $D(\theta)$ be an objective function.  
If there exists a function $G(\theta,\bar{\theta})$ satisfying
\begin{align}
    D(\theta) = \min_{\bar{\theta}} G(\theta,\bar{\theta}),
\end{align}
then $G(\theta,\bar{\theta})$ is called an \emph{auxiliary function} of $D(\theta)$, and $\bar{\theta}$ is referred to as an \emph{auxiliary variable}.
\end{definition}

\begin{theorem}\label{Auxiliary Function Method}
By alternately minimizing the auxiliary function $G(\theta,\bar{\theta})$ with respect to $\bar{\theta}$ and $\theta_i$:
\begin{align}
    &\bar{\theta} \leftarrow \underset{\bar{\theta}}{\arg\min}\, G(\theta,\bar{\theta}),\\
    &\theta_i \leftarrow \underset{\theta_i}{\arg\min}\, G(\theta,\bar{\theta}) \quad (i=1,\ldots,I),
\end{align}
the objective value $D(\theta)$ is guaranteed to decrease monotonically.
\end{theorem}

When constructing an auxiliary function, the following inequality is frequently employed.

\begin{theorem}[Jensen’s Inequality]\label{Jensen's inequality}
For any convex function $g$, real numbers $x_1,\ldots,x_n$, and positive weights $\lambda_1,\ldots,\lambda_n$ satisfying $\sum_{i=1}^{n}\lambda_i = 1$, the following holds:
\begin{align}
    g\!\left(\sum_{i=1}^{n}\lambda_i x_i\right) \le \sum_{i=1}^{n}\lambda_i g(x_i),
\end{align}
with equality if and only if $x_1 = \cdots = x_n$.
\end{theorem}

\section{Kurtosis of GP}\label{Sec: Kurtosis of GP}

In this section, we compare the kurtosis of the generalized Poisson (GP) and negative binomial (NB) distributions.  
Kurtosis serves as a measure of tail heaviness, with larger values indicating heavier tails.

\begin{definition}[Kurtosis]\label{kurtosis}
Let $D$ be a probability distribution, and let $X$ be a random variable following $D$.  
Denote by $\mu$ and $\sigma^{2}$ the mean and variance of $X$, respectively.  
The kurtosis of $D$ is defined as
\begin{align}
	\mathrm{kurtosis} = \frac{E[(X-\mu)^{4}]}{\sigma^{4}} - 3. \label{eq: kurtosis}
\end{align}
The constant 3 is subtracted so that the normal distribution has a kurtosis of zero.
\end{definition}

A positive kurtosis indicates a heavy-tailed distribution, whereas a negative kurtosis corresponds to a light-tailed one.  
For two distributions $D_{1}$ and $D_{2}$ with the same mean and variance, the difference in their kurtosis values reflects the relative heaviness of their tails.  
In what follows, we derive and compare the kurtosis of the GP and NB under the assumption of equal mean and variance.

Let $X \sim \mathrm{GP}(\eta,\mu)$.  
The kurtosis of the GP is given by
\begin{align}
	\mathrm{kurtosis}_{\text{GP}} = \frac{1 + 8\mu + 6\mu^{2}}{(1 - \mu)\eta}, \label{eq: kurtosis of GP}
\end{align}
as reported by \citet{IBEJI2020e00494}.  
When $\mu = 0$, this reduces to $1/\lambda$, which coincides with the kurtosis of the Poisson distribution.  
Using the reparameterization in~\eqref{eq: reparameterization of GP}, we can rewrite~\eqref{eq: kurtosis of GP} as
\begin{align}
	\mathrm{kurtosis}_{\text{GP}} = \frac{1}{\lambda} + \frac{15\theta^{2} + 10\theta}{\lambda}. \label{eq: kurtosis of GP in case of reparameterization}
\end{align}

Next, let $Y \sim \mathrm{NB}(r, p)$.  
The kurtosis of the NB is given by
\begin{align}
	\mathrm{kurtosis}_{\text{NB}} = \frac{6}{r} + \frac{(1 - p)^{2}}{pr}. \label{eq: kurtosis of NB}
\end{align}
By reparameterizing as
\begin{align}
	r = \alpha, \quad p = \frac{\lambda}{\lambda + \alpha},
\end{align}
we obtain
\begin{align}
	\mathrm{kurtosis}_{\text{NB}} = \frac{1}{\lambda} + \frac{6}{\alpha} - \frac{1}{\lambda + \alpha}. \label{eq: kurtosis of NB in case of reparameterization}
\end{align}
Under this parameterization,
\begin{align}
	E[Y] = \lambda, \quad V[Y] = \lambda \left( 1 + \frac{\lambda}{\alpha} \right).
\end{align}
To compare the GP and NB under equal variance, we equate $E[Y] = E[X]$ and solve
\begin{align}
	\lambda \left( 1 + \frac{\lambda}{\alpha} \right) = \lambda (1 + \theta)^{2},
\end{align}
which yields
\begin{align}
	\alpha = \frac{\lambda}{\theta^{2} + 2\theta}.
\end{align}
Substituting this into~\eqref{eq: kurtosis of NB in case of reparameterization} gives
\begin{align}
	\mathrm{kurtosis}_{\text{NB}} = 
	\frac{1}{\lambda} + \frac{6\theta^{4} + 24\theta^{3} + 29\theta^{2} + 10\theta}
	{\lambda(\theta + 1)^{2}}. \label{eq: kurtosis of NB when variance is same as GP}
\end{align}

Finally, subtracting~\eqref{eq: kurtosis of NB when variance is same as GP} from~\eqref{eq: kurtosis of GP in case of reparameterization} yields
\begin{align}
	\mathrm{kurtosis}_{\text{GP}} - \mathrm{kurtosis}_{\text{NB}}
	&= \frac{15\theta^{2} + 10\theta}{\lambda}
	   - \frac{6\theta^{4} + 24\theta^{3} + 29\theta^{2} + 10\theta}
	          {\lambda(\theta + 1)^{2}} \notag\\
	&= \frac{9\theta^{4} + 16\theta^{3} + 6\theta^{2}}
	          {\lambda(\theta + 1)^{2}}. \label{eq: difference of kurtosis}
\end{align}
Since $\lambda > 0$ and $\theta \ge 0$, the right-hand side of~\eqref{eq: difference of kurtosis} is always positive.  
Therefore, the GP distribution is more heavy-tailed than the NB distribution when they share the same mean and variance.

\end{appendices}

\bibliography{sn-bibliography}

\end{document}